\documentclass[lettersize,journal]{IEEEtran}
\usepackage{amsmath,amsfonts}
\usepackage{algorithmic}
\usepackage{algorithm}
\usepackage{array}
\usepackage[caption=false,font=normalsize,labelfont=sf,textfont=sf]{subfig}
\usepackage{textcomp}
\usepackage{stfloats}
\usepackage{url}
\usepackage{verbatim}
\usepackage{multirow}
\usepackage{caption}
\usepackage{ulem} 
\usepackage{soul}
\usepackage{fontawesome}
\usepackage{booktabs}
\usepackage{hyperref}
\usepackage{graphicx}
\usepackage{xcolor}
\usepackage{amsmath}
\usepackage{cite}
\begin{document}

\title{LDGCN: An Edge-End Lightweight Dual GCN Based on Single-Channel EEG for Driver Drowsiness Monitoring}

\author{Jingwei Huang$^{ \dagger}$, Chuansheng Wang$^{ \dagger}$, Jiayan Huang,  \\
Haoyi Fan, Antoni Grau~\IEEEmembership{Senior Member,~IEEE, Fuquan Zhang $^{*}$}

\thanks{

J. Huang is with the College of Computer and Big Data, Fuzhou University, Fuzhou, China. (E-mail: huangjw0824@163.com)

C. Wang and G. Antoni are with the Department of Automatic Control, Polytechnic University of Catalonia, Barcelona, Spain. (E-mail: wangcs95@163.com and antoni.grau@upc.edu)

Jiayan Huang is with the New Engineering Industry College, Putian University, Putian, China. (E-mail: jyan\_huang@163.com)

Haoyi Fan, School of Computer and Artificial Intelligence, Zhengzhou University. (Email: fanhaoyi@zzu.edu.cn)

F. Zhang is with Fujian Provincial Key Laboratory of Information Processing and Intelligent Control, Minjiang University, Fuzhou, China, College of Computer and Control Engineering, Minjiang University, Fuzhou, China, Digital Media Art, Key Laboratory of Sichuan Province, Sichuan Conservatory of Music, Chengdu, China, Fuzhou Technology Innovation Center of Intelligent Manufacturing information System, Minjiang University, Fuzhou, China and Engineering Research Center for Intangible Cultural Heritage (ICH Digitalization and Multi-source Information Fusion (Fujian Polytechnic Normal University), Fujian Province University, Fuzhou, China. (E-mail: 8528750@qq.com)

$^{ \dagger}$ The first two authors contribute equally to this work.
first
$^{*}$ Corresponding authors: Fuquan Zhang.

}

}

\markboth{Journal of \LaTeX\ Class Files,~Vol.~14, No.~8, August~2021}%
{Shell \MakeLowercase{\textit{et al.}}: A Sample Article Using IEEEtran.cls for IEEE Journals}

\IEEEpubid{}

\maketitle

\begin{abstract}
Driver drowsiness electroencephalography (EEG) signal monitoring can timely alert drivers of their drowsiness status, thereby reducing the probability of traffic accidents. Graph convolutional networks (GCNs) have shown significant advancements in processing the non-stationary, time-varying, and non-Euclidean nature of EEG signals. However, the existing single-channel EEG adjacency graph construction process lacks interpretability, which hinders the ability of GCNs to effectively extract adjacency graph features, thus affecting the performance of drowsiness monitoring. To address this issue, we propose an edge-end lightweight dual graph convolutional network (LDGCN). Specifically, we are the first to incorporate neurophysiological knowledge to design a Baseline Drowsiness Status Adjacency Graph (BDSAG), which characterizes driver drowsiness status. Additionally, to express more features within limited EEG data, we introduce the Augmented Graph-level Module (AGM). This module captures global and local information at the graph level, ensuring that BDSAG features remain intact while enhancing effective feature expression capability. Furthermore, to deploy our method on the fourth-generation Raspberry Pi, we utilize Adaptive Pruning Optimization (APO) on both channels and neurons, reducing inference latency by almost half. Experiments on benchmark datasets demonstrate that LDGCN offers the best trade-off between monitoring performance and hardware resource utilization compared to existing state-of-the-art algorithms. All our source code can be found at \url{https://github.com/BryantDom/Driver-Drowsiness-Monitoring}. 
\end{abstract}


\section{INTRODUCTION} \label{sec1:intro}
\begin{figure}
    \centering
    \includegraphics[width=\columnwidth]{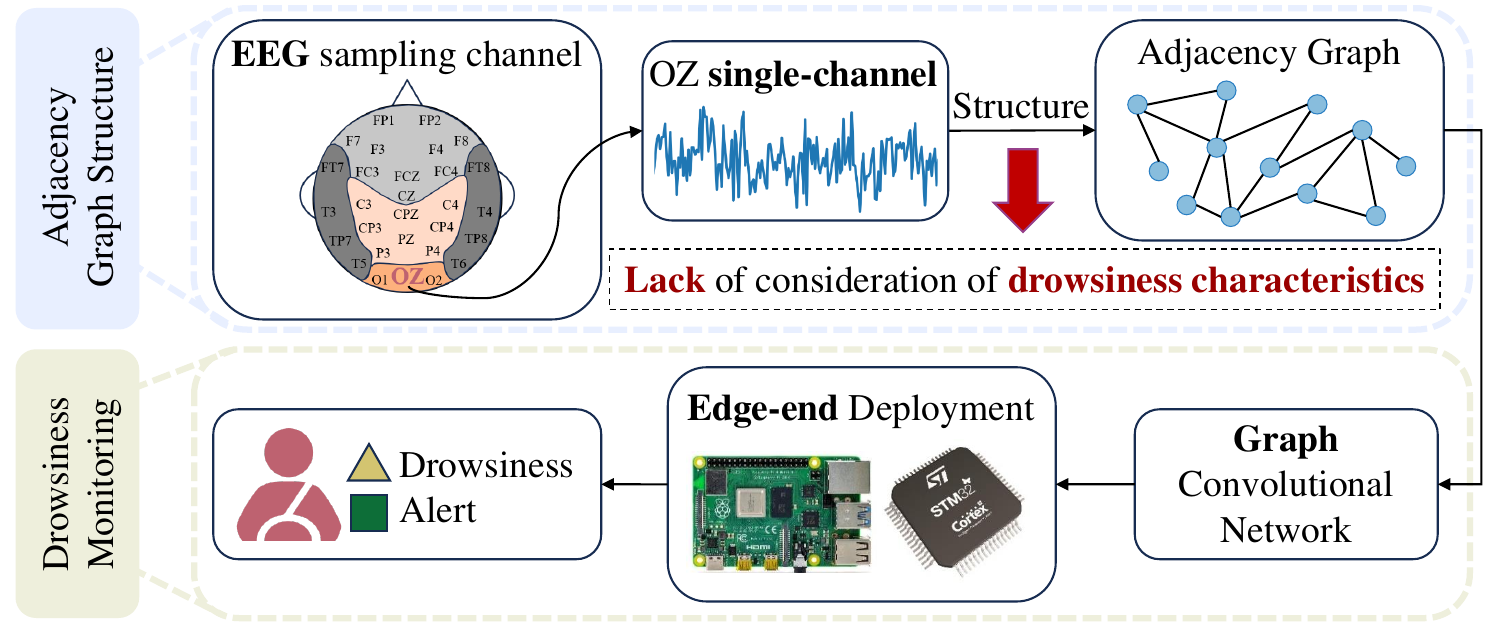}
    \caption{The illustration of our motivation. It first use single-channel EEG (Oz) to construct the adjacency graph for representing its non-Euclidean space information and then use GCN deployed on edge device to identify the driver drowsiness. However, due to existing methods lack of consideration of neurophysiological knowledge on the characteristics of driver drowsiness, it reduces their interpretability and rationality in constructing adjacency graph, which directly affects the recognition accuracy.}
    \label{fig1}
\end{figure}

\IEEEPARstart{F}{atigue} driving leads to inattention, delayed responses, and even a status of drowsiness, posing significant risks to the safety of both drivers and others \cite{i1, t2}. According to statistics, fatigue driving is one of the causes of traffic accidents, accounting for 20\% - 30\% \cite{i2}. The high temporal resolution and low cost of EEG make it has been widely used in non-invasive driver fatigue identification \cite{interpretableCNN}. However, due to the lack of consideration for EEG neurophysiological knowledge of drowsiness characteristics, existing methods struggle to accurately capture the driver's drowsy status. Therefore, establishing a real-time driver drowsiness monitoring method that can be deployed on edge-end devices with high robustness and accuracy remains a challenge, as shown in Fig. \ref{fig1}.

Common EEG-based driver drowsiness monitoring methods are categorized into traditional machine learning-based methods and deep learning-based methods. Typical methods of the former include Random Forest (RF) \cite{RF} and Support Vector Machine (SVM) \cite{svm}. However, these methods heavily rely on manual feature extraction, which can easily lead to significant human errors. Moreover, due to the non-stationary and time-varying characteristics of EEG signals, traditional classifiers have limitations in accuracy. \par

On the contrary, typical deep learning-based methods such as convolutional neural networks (CNNs) \cite{cnn}, Long-Short Term Memory (LSTM) networks \cite{lstm}, and Attention Mechanism (AM) models \cite{am} have demonstrated strong abilities in feature learning for temporal data in extensive studies. For example, Liu et al. \cite{TCACNet} proposed a temporal and channel attention convolutional network (TCACNet) to address the non-stationarity of EEG signals. Wang et al. \cite{tf} used discrete wavelet transform (DWT) as the EEG time-frequency transformation strategy and combined it with an improved CNN to learn frequency domain features. However, such methods typically involve deeper and more complex network structures for hidden feature extraction, significantly increasing both computational complexity and inference time. \par


Since common learning frameworks are typically designed for cloud training, they require a large amount of computational resources, such as those provided by PyTorch, TensorFlow, and Keras. Recently, applying artificial intelligence (AI) technology on edge-end devices has received increasing attention \cite{i10} in academic research and industrial applications. Deployment on edge-end devices avoids personal privacy leakage and significantly reduces the time for transmitting data to the cloud, enabling fast and direct EEG data analysis as well \cite{t1}. However, large-scale deep neural networks face tight computing resource and power consumption limitations when deployed at the edge-end. Therefore, the key issue that needs to be addressed is the trade-off between improving performance and reducing resource consumption. \par

In this work, we propose an edge-end Lightweight Dual GCN for single-channel EEG driver drowsiness monitoring (LDGCN). To our knowledge, our LDGCN is the first to consider the impact of neurophysiological knowledge on EEG drowsiness status recognition. Specifically, we first utilize wavelet transform to perform time-frequency conversion of the EEG signal and construct its corresponding interpretable and reliable Baseline Drowsiness Status Adjacency Graph (BDSAG). Meanwhile, we set up adjacency connectivity coefficients to reduce time complexity and introduce an Augmented Graph-level Module (AGM) to enhance the expression ability of effective features. Subsequently, to overcome the computational limitations of edge devices (the 4th generation Raspberry Pi in this paper), we design a lightweight dual GCN for real-time driver drowsiness monitoring based on single-channel EEG. Finally, we adopt adaptive pruning optimization (APO) to optimize the channels and neurons of the model, thereby reducing latency during the inference phase while improving recognition accuracy. The main contributions of our work are as follows:
\begin{enumerate}
\item{We constructed an interpretable Baseline Drowsiness Status Adjacency Graph (BDSAG) for EEG data, providing a necessary prerequisite for efficient learning of drowsiness features in graph convolution. To our knowledge, we are the first to combine neurophysiological knowledge to construct an adjacency graph.}
\item{We proposed an edge-end based lightweight dual GCN (LDGCN) for real-time driver drowsiness monitoring, achieving a good trade-off between recognition accuracy and resource usage. Moreover, the LDGCN can successfully capture both global and local views to enhance the expression ability of effective features within the limited graph level data.}
\item{We reduced the inference delay by approximately half (about 44.44 ms) and improved the accuracy by 0.8\% by using adaptive pruning optimization (APO) for both channels and neurons.}
\item{Extensive experiments on the publicly available benchmark drowsiness dataset show that our LDGCN achieves higher recognition performance and lower resource consumption than state-of-the-art models.}
\end{enumerate}

The rest sections of this article are organized as follows. The Section \ref{sec2:related} reviews the related works. The Section \ref{sec3:method} introduces the designed details of the LDGCN. The Section \ref{sec4:experiments} conducts and analyses experiments. The Section \ref{sec5:conclusion} provides a summary of this article.

\section{RELATED WORKS}\label{sec2:related}

In this section, we will respectively introduce neurophysiological knowledge about driver drowsiness characteristics of EEG signals, EEG data augmentation methods, and some common pruning strategies and corresponding potential challenges.

\subsection{Neurophysiological Knowledge of Drowsiness EEG Signals}\label{sec:2-1}
EEG records brain activity by measuring voltage fluctuations generated by ion currents within brain neurons. Jian et al. \cite{EEG-based} have demonstrated that there is a strong correlation between EEG and driver drowsiness status. EEG offers the advantage of directly monitoring brain activity with high temporal resolution, allowing for the early identification of drowsiness status \cite{i8}. Additionally, neurophysiological studies indicate a close relationship between the frequency pattern of EEG signals and drowsiness status. Typically, EEG signals are divided into four frequency patterns: Delta ($\delta$, 0-4 Hz), Theta ($\theta$, 4-8 Hz), Alpha ($\alpha$, 8-12 Hz), and Beta ($\beta$, 12-20 Hz). Research has shown that the power of the $\theta$ and $\alpha$ frequency bands significantly increases when drivers are drowsy \cite{EEG-based, i12}. Motivated by this finding, we constructed the BDSAG incorporating neurophysiological knowledge of drowsiness characteristics, particularly the abnormal changes in the $\theta$ and $\alpha$ frequency bands, to ensure the biological interpretability of adjacency graphs and enhance the analysis and expression of effective features.
\begin{figure*}
    \centering
    \includegraphics[width=\textwidth]{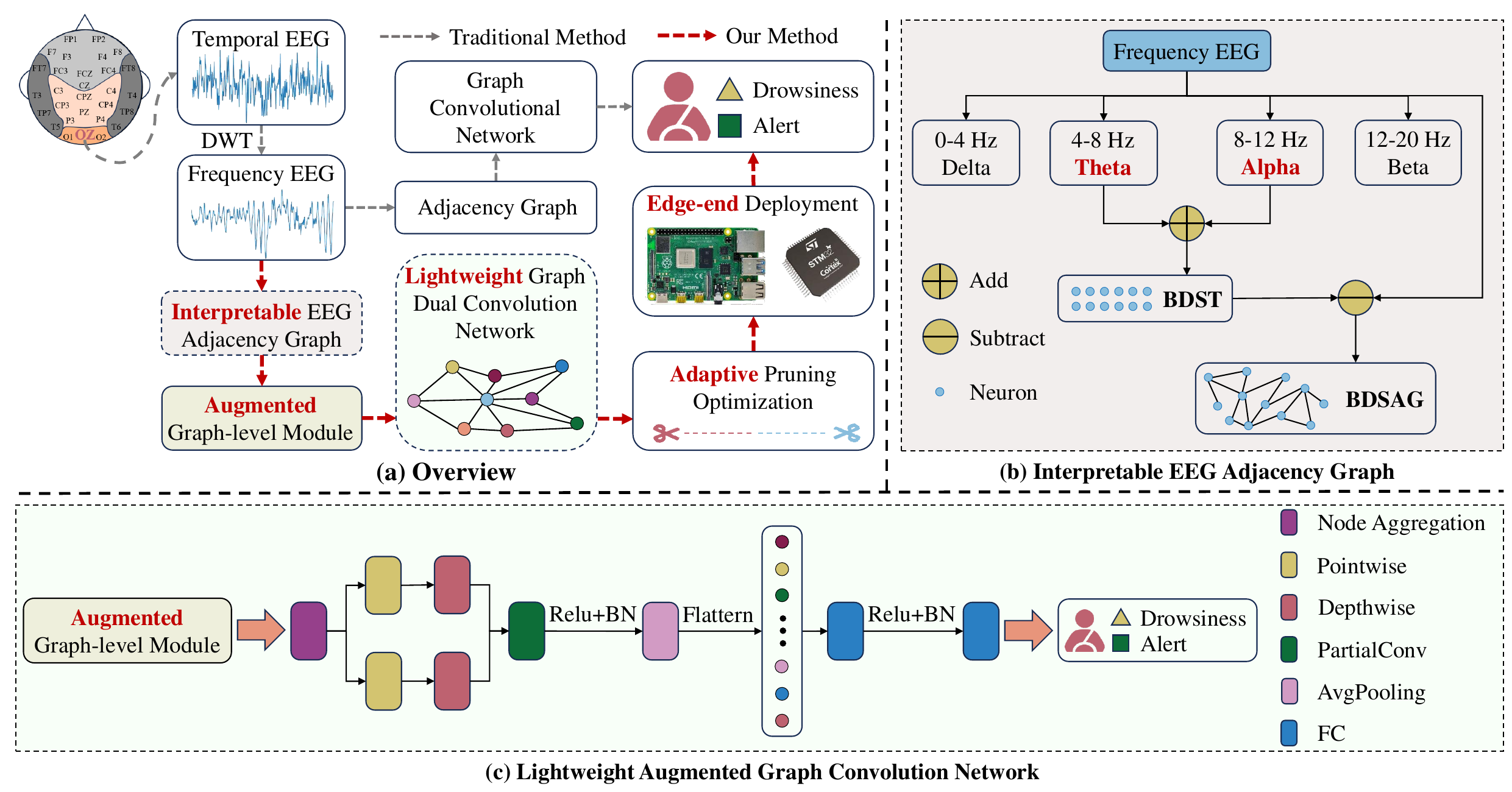}
    \caption{The overview of our drowsiness EEG signal monitoring.}
    \label{fig2}
\end{figure*}
\subsection{Driver Drowsiness Monitoring by Graph Convolutional Network}\label{sec:2-2}
In the EEG classification task, the GCN typically utilizes the topological structure between nodes to aggregate potential features of neighboring samples, and its effectiveness has been widely verified. Tang et al. \cite{r11} constructed an adjacency graph based on the correlation between electrode spatial and spectral features, using diffusion convolution to learn features from each channel. However, this approach is not suitable for single-channel EEG signals, which require rapid processing. Sun et al. \cite{eeg-arnn} used multiple same-temporary feature extraction modules (TFEM) and channel active reasoning modules (CARM) to learn features from each graph structure, but this method suffers from high computational complexity and significant time consumption. Zhuang et al. \cite{cagnn} introduced a squeeze-and-excitation (SE) block to extract the interdependencies of EEG features and utilized an auto-encoder to further learn hidden features. However, the SE block overlooks fluctuations in the frequency domain, compromising the integrity of effective information. Wang et al. \cite{ssgcnet} designed a sparse spectra graph convolutional network (SSGCNet) that effectively reduces running time by removing redundant edges in the adjacency graph. Nonetheless, the single-layer traditional convolution used in SSGCNet is easily affected by the method of constructing the adjacency graph, resulting in low feature fault tolerance and robustness. In contrast, our Lightweight Graph Dual Convolution Network (LGDCN) ensures recognition accuracy while maintaining low resource consumption.

\subsection{Model Pruning Strategies}\label{sec:2-3}
The key to deploying models on resource-limited edge devices is reducing computational costs \cite{t3}. Therefore, balancing reduced computational complexity with maintained recognition performance presents a significant challenge. Pruning can lower computational costs by removing unnecessary connections from the model. Frankle et al. \cite{r3} discovered that pruning could actually enhance accuracy. Lin et al. \cite{r4} implemented a uniform L1 norm-based pruning method across all layers, which may lead to suboptimization. Sun et al. \cite{r5} used a scaling factor to gauge the importance of parameters, yet overlooked the redundant information generated by outputs. Xiang et al. \cite{tucker} combined pruning with tensor decomposition to compress CNNs, which, while effective, disrupts the original model structure and complicates deployment. To address this issue, group-Lasso regularization methods were proposed \cite{r14}. This set all parameters of the pruned filter to zero, effectively removing them mathematically. However, such operations can only reduce the parameter scale to a certain extent. In contrast, our approach involves pruning from both channel and neuron perspectives during the model inference stage to optimize accuracy and reduce inference latency, providing significant application value for edge-end device deployment.

\section{METHOD}\label{sec3:method}
In this section, we present our edge-end based lightweight dual GCN (LDGCN) for driver drowsiness EEG signal monitoring.

\subsection{Preliminary}
Before introduce our LDGCN, we define the concept of related parameters in this work.

    \subsubsection{Time domain of EEG signal} The signal-channel EEG signal $\mathrm{z}=\{\mathrm{z}_1,\mathrm{z}_2,\mathrm{z}_3, \cdots ,\mathrm{z}_n\}$, where $\mathrm{z}_i$ denotes the $i$-th sampling point, and $n$ is the total number of EEG sampling point.
    
    \subsubsection{Frequency domain of EEG signal} Defined as $\mathrm{x}=\{\mathrm{x}_1,\mathrm{x}_2,\mathrm{x}_3, \cdots ,\mathrm{x}_n\}$, where $\mathrm{x}_i$ denotes the corresponding $i$-th frequency signal of the $\mathrm{z}$, and the $\mathrm{x}$ can be calculated by,
    \begin{equation}
    \mathrm{x}=\sum_{i=1}^{n}\mathrm{z}_{i}\cdot \Psi_{i} \hspace{1ex}, \quad i \in [1,2,3,\cdots,n]   \label{eq:p1}
    \end{equation}
    where $\Psi_{i}$ is the basis function of wavelet transform for the time-frequency conversion task.
    
    \subsubsection{The graph representation of EEG signal}The definition of a graph is $\mathcal{G} = (\mathcal{V},\mathcal{E})$, where $\mathcal{V}=\{v_1,v_2,v_3, \cdots,v_n\}$ denotes the node set. The node $v_i$ corresponds to the EEG sampling point $\mathrm{x}_i$. $\mathcal{E}$ denotes the edge set, which is defined as,
    \begin{equation}
    \mathcal{E} = \begin{bmatrix}
      &E_{1,1}  &E_{1,2}  &\cdots   & E_{1,n} \\
      &E_{1,2}  &E_{2,2}  &\cdots  &E_{2,n} \\
      &\cdots  &\cdots  &\cdots  &\cdots \\
      &E_{1,n}  &E_{2,n}  &\cdots  &E_{n,n}
    \end{bmatrix} \label{eq:e}
    \end{equation}
    where each $E_{i,j}$ denotes the feature adjacency between nodes $v_i$ and $v_j$.
\subsection{Construction of Interpretable EEG Adjacency Graph}
Owing to the important EEG features contained in the adjacency graph (AG), it is the prerequisite to ensure the efficiency of graph convolution. However, due to most existing AGs lack consideration of neurophysiological knowledge, which leads to its insufficient interpreability and classification reliability. Therefore, our objective is to aim to construct the interpretable EEG AG. As shown in Fig. \ref{fig2} (b), the process is mainly to select Baseline Drowsiness Status Tensor (BDST) and construct Baseline Drowsiness Status Adjacency Graph (BDSAG).

\begin{figure}
  \captionsetup{justification=raggedright, singlelinecheck=false}
  \centering
  \renewcommand{\figurename}{Fig.}
  \includegraphics[width=0.49\textwidth]{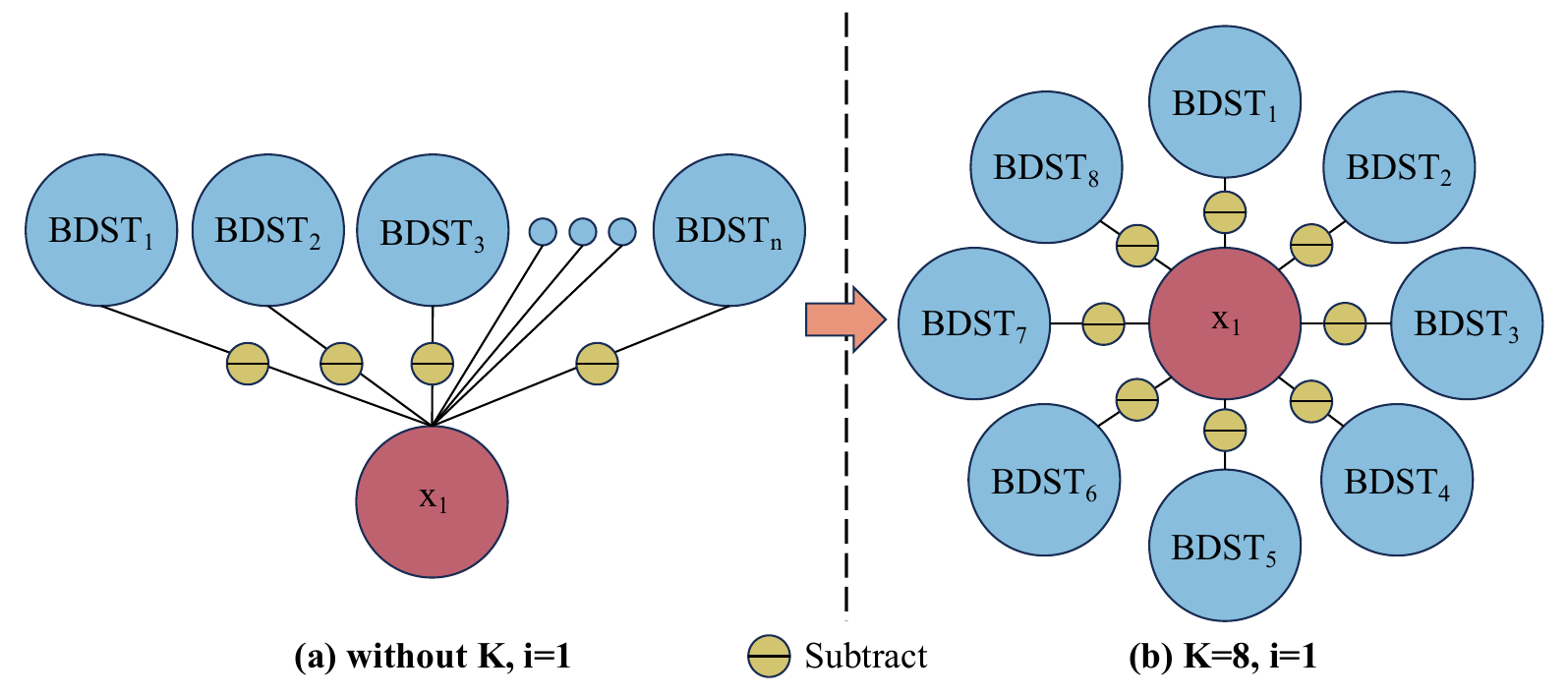}
  \caption{Schematic representation of ${\mathrm{x}_1}$ and BDST computation process. (a) Time complexity is $O(n^2)$ without adjacency connectivity coefficient K. (b) adjacency connectivity coefficient K = 8, the time complexity is $O(n)$.}
  \label{fig3}
\end{figure}

\subsubsection{Baseline Drowsiness Status Tensor (BDST)}
Firstly, we extract the four different frequency band features from EEG signal by filtering and separate it in frequency domain, including $\sigma$, $\theta$, $\alpha$, and $\beta$. According to the observation that the power of $\theta$ and $\alpha$ will significantly increase when drivers in drowsiness status \cite{i12}, we decide to select the average value of both of them as BDST for retaining the key frequency domain features of drowsiness, which can be expressed as,
\begin{equation}
    BDST = \sum_{i=1}^{n} \frac{({\theta}_{i}+{\alpha}_{i})}{2}  \hspace{1ex}, \quad i \in [1,2,3,\cdots,n] \label{t-a2}
\end{equation}

\subsubsection{Baseline Drowsiness Status Adjacency Graph (BDSAG)}
After obtaining the BDST, it is critical to identify the feature adjacency relationships between sampling points and construct the BDSAG. Previous methods typically calculated their weighted edge value based on the change difference between two sampling points. However, such methods have the two disadvantages. On the one hand, since they tend to calculate their weight value based on distance, the interpretability of weights is insufficient. On the other hand, the high computational complexity may lead to low efficiency in practical applications. On this basis, we take into account the drowsiness features. In other words, when the $r$-th raw value of EEG frequency domain signal $\mathrm{x}_r$ approaches the $BDST_c$ (the $c$-th column value of BDST), it indicates a higher probability of drowsiness. On the contrary, it may be in a wakefulness when the difference between them is far apart. Notably, due to the fact that the weight values between $\mathrm{x}_r$ and $BDST_c$ are the same, the constructed graph is a symmetric weigthed undirected graph, which can be defined as,
\begin{equation}
E_{r,c}, E_{c,r} = 
\left\{
\begin{array}{ll}
\frac{\sum_{r = 1}^{n}\sum_{c = 1}^{n} \mathrm{x}_r-BDST_{c} }{\left | r-c \right | }\hspace{1ex}, & \text{if } r \ne c \\
1\hspace{1ex}, & \text{if } r = c  \label{ag-1}
\end{array}
\right.
\end{equation}
where $E_{r,c}$, $E_{c,r}$ denote the weight relationship between $\mathrm{x}_r$ and $BDST_c$, and the column index $r, c \in [1, 2, 3, \cdots, n]$.

However, we found that this AG construction method has a high time complexity ($O(n^2)$). Therefore, we introduce an additional adjacency connectivity coefficient $K$ to specify the range of column calculations, rather than being limited by the total number of sampling points $n$. In this work, through repeated experiments, we found that the time complexity can be reduced to $O(n)$ when the $K$ value ranging between 5 and 10. In particular, to more intuitively demonstrate the effect of $K$, we visualized it in Fig. \ref{fig3}. Meanwhile, we modified Eq. \ref{ag-1} as,
\begin{equation}
E_{r,c}, E_{c,r} = 
\left\{
\begin{array}{ll}
\frac{\sum_{r = 1}^{n}\sum_{c = r}^{r+K} \mathrm{x}_r-BDST_{c} }{\left | r-c \right | }\hspace{1ex}, & \text{if } r \ne c \\
1\hspace{1ex}, & \text{if } r = c  \label{ag-2}
\end{array}
\right.
\end{equation}

\subsection{Augmented Graph-level Module (AGM)}
Based on the constructed BDSAG, we can convert EEG data into graph dataset, which can augment the limited dataset based at graph level. Existing methods typically augment their data based on the time series (one-dimensional direction) of EEG signals. However, such method may destroy the structure features of the graph. Additionally, it has been verified that sampling for graph subsets can benefit downstream tasks on various graph datasets \cite{Lggnet}. Inspired by that, we tend to focus on the AG itself and achieve data augmentation based on the graph level in this work. Specifically, we use larger and smaller sub-graphs to represent the global and local view respectively, where the global view can be expressed as, 
\begin{equation}
\mathcal G_g = f(\mathcal G) \cdot R \hspace{1ex}, \quad R \in [0.5, 1]
\end{equation}
where $f(\cdot)$ denotes the sampling method of view, $R$ is the ratio of retained graph level feature. Similarly, we express the local view as,
\begin{equation}
\mathcal G_l = f(\mathcal G) \cdot r \hspace{1ex}, \quad R \in [0, 0.5]
\end{equation}
where $r$ is the ratio of retained graph level feature. So far, we finally combine BDSAG with the global and local views, which can be defined as, 
\begin{equation}
\tilde{\mathcal G} = g({\mathcal G|\mathcal G}_g|{\mathcal G}_l)  \label{glc}
\end{equation}
where $\tilde{\mathcal G}$ denotes the augmented graph. Additionally, to improve the data stability and the model's representative ability for complex EEG feature, we also introduce a nonlinear change function $g(\cdot)$ to map the features into the latent space. 

\subsection{The Structure of LDGCN}
Obviously, compared to one-dimensional information, the graph structure contains richer and more complete features. Meanwhile, the computational complexity of GCN needs to be minimized for resource-limited edge device to ensure the model efficiency and deployability. For this, we design our lightweight dual graph convolutional network (LDGCN) for drowsiness EEG monitoring.\par

The structure of LDGCN as shown in Fig. \ref{fig2} (c). Firstly, LDGCN performs aggregation operations on each node of the AGM enhanced graph level data to extract features from adjacent nodes and achieve feature dimensionality reduction. This step is considered one of the key steps in graph convolution.\par

Then, considering that depthwise separable convolution has better computational complexity than traditional convolution \cite {dsc}. In addition, it can better ensure its integrity during feature interaction, which has a natural advantage in efficiently extracting graph structure features. Specifically, depthwise separable convolution consists of pointwise convolution and depthwise convolution. Pointwise convolution does not involve the mixing of spatial information when performing linear transformations between channels. Therefore, although it will change the relationship between channels, it will not affect the spatial structure within the channels. Depthwise convolution is processed within independent spatial regions, and the positional relationships of elements within the channel remain unchanged, thus maintaining the spatial structure within the channel. This enables the network to learn more accurate representation of graph features. Importantly, we differ from classical depthwise separable convolutions in the order of operations \cite{Xception}. Performing pointwise convolution initially allows the decomposition of the convolution operation on the channel dimension into separate convolutions across multiple channels. This significantly reduces the computational cost of depthwise convolution, making it highly advantageous for edge-end or resource-limited scenarios. Furthermore, we employ a dual convolutional structure to convey and interact information from various perspectives or at different levels, thereby enhancing the model’s robustness to a certain extent. \par

Next, we establish a fusion layer, composed of partial convolution (PartialConv) as described in \cite{pconv}. This approach allows us to control the feature flow autonomously, selectively transferring certain input features to the subsequent layer while maintaining the features of other parts unchanged. This not only curtails the propagation of irrelevant information but also renders the network more compact. Following this, we employ a Relu activation layer and a BatchNorm (BN) layer to prevent the gradient vanishing phenomenon induced by convolution operations. Subsequently, we incorporate a global average pooling (AvgPooling) layer to significantly reduce the number of parameters. \par

Finally, with regard to the establishment of the fully connected (FC) layer, it serves to augment the non-linear capability of the model. This enables the model to more effectively capture the semantic features of the data, thereby enhancing the accuracy of classification tasks. \par

\subsection{Adaptive Pruning Optimization for Edge-end Deployment}
In this section, we introduce the adaptive optimization optimization (APO) in section \ref{sec:adp} and the technical solutions for edge-end deployment in section \ref{sec:eed}.

\subsubsection{Adaptive Pruning Optimization (APO)}\label{sec:adp}
\begin{figure}
  \captionsetup{justification=raggedright,singlelinecheck=false}
  \centering
  \renewcommand{\figurename}{Fig.}
  \includegraphics[width=0.49\textwidth]{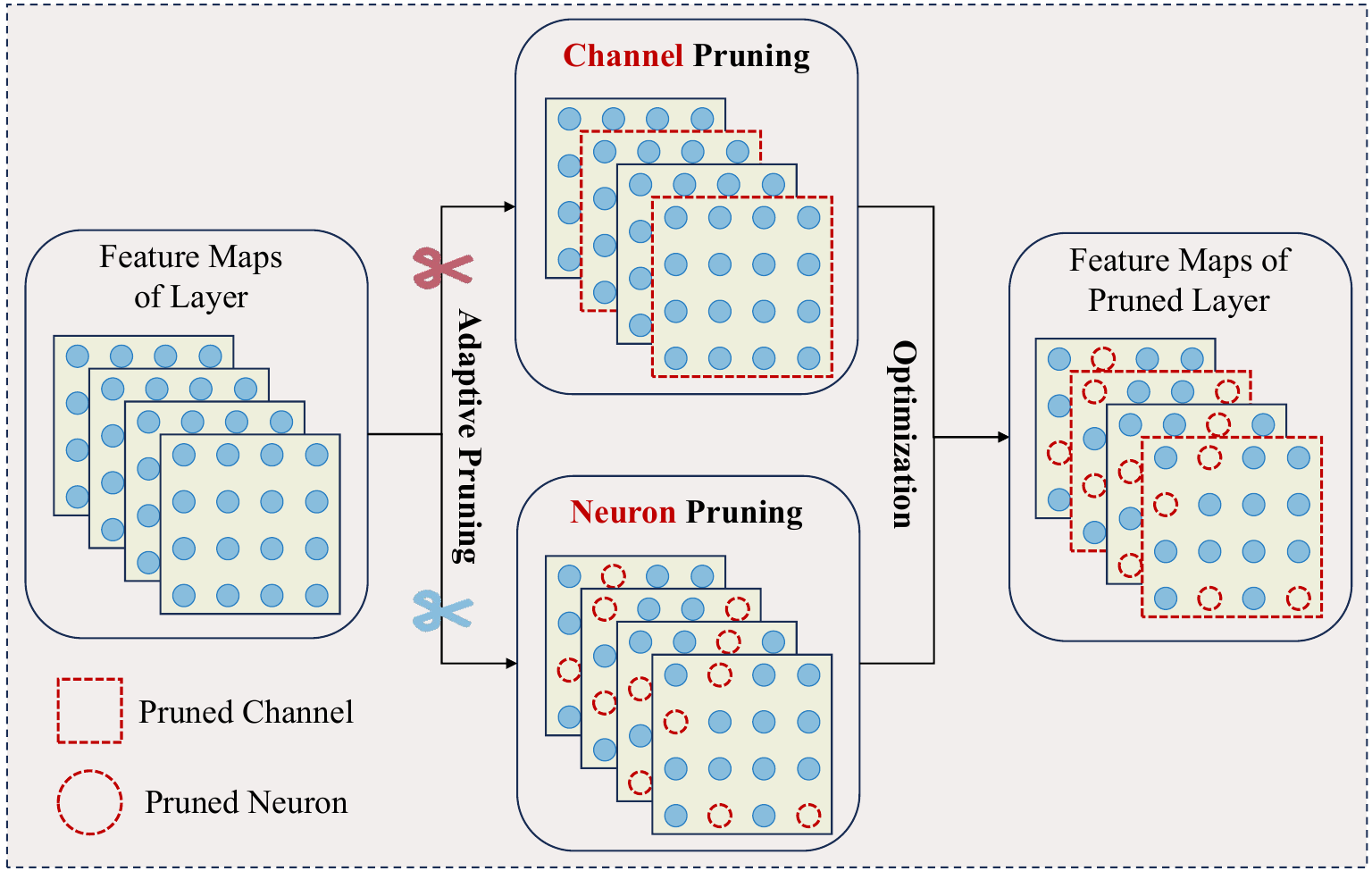}
  \caption{The adaptive channel-neuron pruning optimization of LDGCN.}
  \label{fig4}
\end{figure}

To deploy the model on an edge-end device efficiently, we propose an adaptive pruning optimization (APO) strategy which can improve the model accuracy and significantly reduce the inference delay. In fact, the parameter number directly affects model's computational complexity, and model pruning aims to reduce the parameter scale. During the experimental process, we noticed a fluctuation in the final recognition accuracy when the output channel of the convolution was set to either 4 or 8, or the neuron number of full connected layer is set to 128 or 256. Therefore, we speculate that there may be an optimal parameter reduction configuration for time reducing and accuracy improvement when output channel in [4, 8] and neuron number in [128, 256]. And we have proved this conjecture in Table \ref{tab:pruning_compare}. Based on the above validation, we have devised an APO that considers two aspects, as shown in Fig. \ref{fig4}. 

On the one hand, a channel pruning optimization for convolutional layers is designed to filter the channel weight value. For unimportant channels $P_{cindex}$, the filtering calculation method is as follows,
\begin{equation}
P_{cindex}=argminsort(\sum_{i=1}^{channel}|w_i| )[channel\cdot PR] \label{eq:pruning_channel}
\end{equation}
where $argminsort$ represents the importance sorting of channels in ascending order, $w_i$ is the weight in the channel, $channel$ is the total channel number in current layer, and $PR$ is the pruning ratio. and $channel\cdot PR$ denotes the number of pruning channel.

On the other hand, a neuron pruning optimization is designed to filter the neuron weight value. Similarly, for unimportant neurons $P_{nindex}$, the filtering calculation method is as follows,
\begin{equation}
P_{nindex}=argmidsort(\sum_{i=1}^{neuron}|w_i|)[neuron\cdot pr] \label{eq:pruning_neuron}
\end{equation}
where $argmidsort$ represents the importance sorting of neurons in median-based order. This mainly because there is a lot of redundant information in the fully connected layer, and median-based order can discard unimportant features as much as possible while improving the expressiveness. $neuron$ is the total number of neurons in current channel, and $pr$ is the pruning ratio. 

Subsequently, we set the weights of $P_{cindex}$ and $P_{nindex}$ to 0, while keeping the other weights unchanged. This optimization method makes the LDGCN sparse, and thus it can reduce the computation delay time.

\subsubsection{Edge-end Deployment} \label{sec:eed}
\begin{figure}
  \captionsetup{justification=raggedright,singlelinecheck=false}
  \centering
  \renewcommand{\figurename}{Fig.}
  \includegraphics[width=0.49\textwidth]{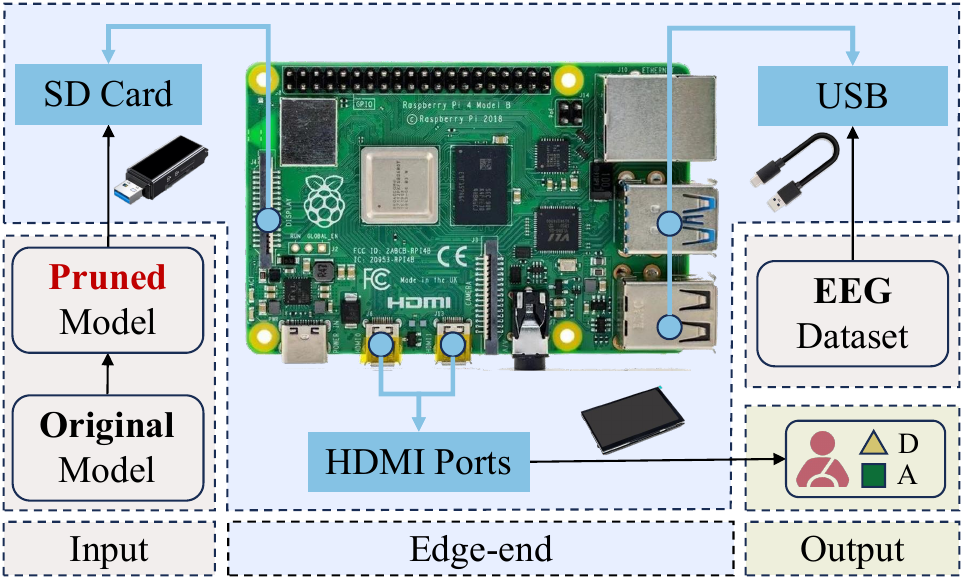}
  \caption{Framework diagram for pruned model deployment edge-end.}
  \label{fig5}
\end{figure}

Due to the complex structures and high computational requirements of deep learning models, deployment on resource-limited edge-end device becomes a challenge. Deploying deep learning models at the edge-end obviates the need for transferring all data to the cloud for processing and analysis. This not only safeguards user privacy but also reduces the time consumed in data transfer. Such an approach is particularly vital for real-time monitoring of driver drowsiness. To showcase the practical applicability of our approach, we deploy the pruned LDGCN to the edge-end devices during the inference phase. As shown in \ref{fig5}, we store the pruned model in SD card through read-write device and input the EEG dataset through USB interface for achieving edge-end analysis. Subsequently, we output the final drowsiness recognition results (including D: Drowsiness, A: Alert) the display through HDMI ports.

\section{EXPERIMENTS}\label{sec4:experiments}
In this section, we assess the effectiveness of our method on a publicly and available benchmarking dataset, which is specifically used for monitoring driver drowsiness \cite{single-eeg}. We conduct experiments on model recognition performance and the ablation studies of our proposed BDSAG, AGM, APO, and the network structure of our LDGCN.

\subsection{Datasets} 
In this work, we use the publicly and available benchmarking dataset \cite{single-eeg} to evaluate the driver drowsiness monitoring performance of our LDGCN. The dataset contains 2022 samples from 11 different subjects, including 1011 drowsy status (Drowsiness) samples and an equal number of awake status (Alert) samples. All EEG signal samples are captured by using 30 electrodes at a sampling frequency of 500 Hz for participating drivers, who are engaged in a continuous driving task within a virtual reality simulator. The sample distribution for each subject is detailed in Table \ref{dataset}. Each sample measures 30 (channels) × 384 (sample points) in size. Prior research has suggested the feasibility of using only the single-channel ‘Oz’, which is most relevant to the drowsy status, for drowsiness monitoring \cite{single-eeg2}.

\begin{table}[h]
\caption{Drowsiness and Alert sample numbers of 11 drivers in the dataset.}
\renewcommand{\figurename}{Fig.}
\centering
\label{dataset}
\resizebox{0.49\textwidth}{!}{
\begin{tabular}{cc|ccccccccccc|c}
\hline
\multicolumn{2}{c|}{Subject ID}                                                                             & 1  & 2  & 3  & 4  & 5   & 6  & 7  & 8   & 9   & 10 & 11  & Total \\ \hline
\multicolumn{1}{c|}{\multirow{2}{*}{\begin{tabular}[c]{@{}c@{}}Sample\\  Number\end{tabular}}} & Alert      & 94 & 66 & 75 & 74 & 112 & 83 & 51 & 132 & 157 & 54 & 113 & 1011  \\ \cline{2-2}
\multicolumn{1}{c|}{}                                                                          & Drowsiness & 94 & 66 & 75 & 74 & 112 & 83 & 51 & 132 & 157 & 54 & 113 & 1011  \\ \hline
\end{tabular}
}
\end{table}

\subsection{Evaluation Metrics}
To better guide the trade-off between performance and resources in our LDGCN, we evaluate it using both performance metrics and resource metrics.

\subsubsection{Performance Metrics}
We assess our model's performance with five key metrics \cite{EEG-based}, including Accuracy (Acc.), Precision (Pre.), Recall (Rec.), Specificity (Spe.), and F1 Score (F1). Accuracy measures the model's capability to correctly identify the driver's drowsiness or alert status. Precision quantifies how accurately the model identifies drivers as drowsy. Recall evaluates the model's proficiency in recognizing actual drowsy drivers, while Specificity assesses its ability to correctly identify drivers who are truly alert. The F1 Score combines precision and recall, offering a comprehensive metric for evaluating overall model performance.

\subsubsection{Resource Metrics}
Similarly, we measure our model’s resource consumption using five metrics \cite{On-device}, including Footprint, Million Floating Point Operations Per Second (MFLOPs), Latency, Power, and Energy. Footprint measures the size of the data or program in memory, MFLOPs assess the system's speed in handling floating point operations, Latency refers to the time taken to complete a task, Power gauges the amount of electrical energy the system consumes, and Energy represents the total energy consumption of the system over a specific period.

\subsection{Implementation Details}
Our experimental implementation encompasses both training and deployment procedures.

\subsubsection{Training Procedure}
The experiments were conducted on a single GPU, an Nvidia GTX 3060 12GB, utilizing the PyTorch framework. We trained the proposed LDGCN using the Adam optimizer over 30 epochs with a learning rate of 0.0015 and a batch size of 32. The Negative Log-Likelihood (NLL) was employed as the loss function.

\subsubsection{Deployment Procedure}
In this work, we effectively perform real-time driver drowsiness monitoring on edge-end devices, specifically the fourth generation Raspberry Pi (4GB RAM, 16GB Flash). The widespread use and active community surrounding the fourth generation Raspberry Pi facilitate the reproduction of our work and highlight its popularity for local real-time development tasks on edge devices. For deployment, we installed the Linux-based Raspberry Pi OS and the necessary dependencies, successfully executing the inference phase of our LDGCN on this device.

\begin{table*}[t]
\caption{The drowsiness monitoring performance comparison of different methods on 11 different drivers. The best and second-best results are marked in red and blue, respectively.}
\renewcommand{\figurename}{Fig.}
\centering
\label{tab:perfomance_11}
\resizebox{0.99\textwidth}{!}{
\begin{tabular}{cccccccccccccccccccccc}
\specialrule{0.2em}{0.1em}{0.1em}
                                         & \multicolumn{3}{c}{EEG-ARNN \hspace{0.3ex} \cite{eeg-arnn}} & \multicolumn{3}{c}{FBMSNet \hspace{0.3ex}\cite{FBMSNet}}                           & \multicolumn{3}{c}{LGGNet\hspace{0.3ex} \cite{Lggnet}}                            & \multicolumn{3}{c}{SSGCNet \hspace{0.3ex}\cite{ssgcnet}}                                                                                           & \multicolumn{3}{c}{CAGNN \hspace{0.3ex} \cite{cagnn}}                             & \multicolumn{3}{c}{TSception \hspace{0.3ex} \cite{tsception}}                         & \multicolumn{3}{c}{Ours}                                                                                            \\  \cline{2-22}  
\multirow{-2}{*}{ID} & Acc.     & Pre.    & Spe.    & Acc.                                  & Pre.  & Spe.  & Acc.                                  & Pre.  & Spe.  & Acc.                                  & Pre.                                  & Spe.                                  & Acc.                                  & Pre.  & Spe.  & Acc.                                  & Pre.  & Spe.  & Acc.                                  & Pre.                                  & Spe.                                  \\ \hline
1                    & 69.14    & 67.30   & 63.82   & 76.59                                 & 85.71 & 89.36 & 79.78                                 & 86.84 & 89.36 & {\color[HTML]{9A0000} \textbf{81.38}} & 83.90                                 & 85.10                                 & 76.13                                 & 77.48 & 87.23 & 78.72                                 & 76.47 & 74.47 & {\color[HTML]{3531FF} \textbf{80.85}} & 87.17                                 & 89.36                                 \\
2                    & 47.72    & 60.64   & 61.87   & {\color[HTML]{9A0000} \textbf{59.84}} & 69.81 & 69.69 & 50.75                                 & 50.58 & 36.36 & 56.06                                 & 54.44                                 & 37.87                                 & 51.52                                 & 51.62 & 59.15 & {\color[HTML]{3531FF} \textbf{56.82}} & 54.74 & 34.85 & 56.81                                 & 55.99                                 & 49.99                                 \\
3                    & 55.33    & 58.03   & 54.46   & 54.54                                 & 70.17 & 78.29 & 53.33                                 & 72.72 & 96.05 & {\color[HTML]{3531FF} \textbf{64.00}} & 69.81                                 & 78.66                                 & 54.05                                 & 66.14 & 10.76 & 55.33                                 & 78.57 & 96.00 & {\color[HTML]{9A0000} \textbf{64.66}} & 62.22                                 & 54.66                                 \\
4                    & 60.81    & 58.07   & 52.10   & {\color[HTML]{9A0000} \textbf{81.75}} & 78.64 & 73.38 & 72.97                                 & 72.36 & 71.62 & 56.08                                 & 63.63                                 & 83.78                                 & 76.47                                 & 76.58 & 79.72 & 77.70                                 & 69.52 & 56.76 & {\color[HTML]{3531FF} \textbf{80.40}} & 78.48                                 & 77.02                                 \\
5                    & 57.14    & 57.98   & 54.39   & 63.39                                 & 75.88 & 83.84 & 71.42                                 & 94.44 & 97.32 & 68.75                                 & 95.65                                 & 98.21                                 & 68.85                                 & 79.79 & 38.45 & {\color[HTML]{3531FF} \textbf{78.57}} & 94.44 & 96.43 & {\color[HTML]{9A0000} \textbf{81.69}} & 83.80                                 & 84.82                                 \\
6                    & 63.85    & 59.63   & 60.11   & 72.28                                 & 77.15 & 84.72 & 81.92                                 & 93.44 & 95.18 & {\color[HTML]{9A0000} \textbf{84.33}} & 83.13                                 & 83.52                                 & 77.14                                 & 82.62 & 56.65 & 77.71                                 & 87.10 & 90.36 & {\color[HTML]{3531FF} \textbf{83.13}} & 87.67                                 & 89.15                                 \\
7                    & 55.88    & 59.40   & 60.36   & 50.13                                 & 71.94 & 78.91 & {\color[HTML]{3531FF} \textbf{60.72}} & 57.97 & 43.13 & 55.88                                 & 87.49                                 & 98.03                                 & 57.81                                 & 59.15 & 76.57 & 52.94                                 & 51.95 & 27.45 & {\color[HTML]{9A0000} \textbf{69.60}} & 69.23                                 & 68.62                                 \\
8                    & 58.71    & 59.24   & 59.67   & 54.16                                 & 64.52 & 65.50 & 59.84                                 & 55.50 & 20.45 & 57.95                                 & 54.83                                 & 25.75                                 & {\color[HTML]{3531FF} \textbf{62.17}} & 77.08 & 99.25 & 53.79                                 & 51.97 & 27.85 & {\color[HTML]{9A0000} \textbf{77.27}} & 76.08                                 & 74.99                                 \\
9                    & 64.01    & 59.85   & 59.24   & 85.03                                 & 67.63 & 66.94 & {\color[HTML]{3531FF} \textbf{88.85}} & 84.65 & 82.80 & 79.93                                 & 73.97                                 & 67.51                                 & 72.63                                 & 78.69 & 95.50 & 76.11                                 & 67.83 & 52.87 & {\color[HTML]{9A0000} \textbf{92.99}} & 96.55                                 & 96.81                                 \\
10                   & 74.07    & 60.96   & 61.35   & 82.40                                 & 68.77 & 68.59 & 85.18                                 & 97.52 & 98.14 & {\color[HTML]{3531FF} \textbf{87.03}} & 91.66                                 & 92.59                                 & 76.92                                 & 81.62 & 57.42 & 85.19                                 & 93.18 & 94.44 & {\color[HTML]{9A0000} \textbf{88.88}} & 97.61                                 & 98.14                                 \\
11                   & 66.37    & 61.76   & 62.71   & 61.94                                 & 69.50 & 71.71 & 69.02                                 & 75.29 & 81.41 & {\color[HTML]{9A0000} \textbf{77.43}} & 83.69                                 & 86.72                                 & 61.59                                 & 75.06 & 24.86 & 66.81                                 & 83.93 & 92.04 & {\color[HTML]{3531FF} \textbf{76.10}} & 77.57                                 & 78.76                                 \\ \hline
Ave.                 & 61.18    & 60.26   & 59.10   & 67.40                                 & 72.22 & 76.04 & 70.35                                 & 75.48 & 73.80 & {\color[HTML]{3531FF} \textbf{71.53}} & {\color[HTML]{3531FF} \textbf{76.60}} & {\color[HTML]{3531FF} \textbf{76.12}} & 66.79                                 & 73.19 & 62.28 & 69.06                                 & 73.60 & 65.74 & {\color[HTML]{9A0000} \textbf{77.76}} & {\color[HTML]{9A0000} \textbf{79.97}} & {\color[HTML]{9A0000} \textbf{79.26}} \\ \specialrule{0.2em}{0.1em}{0.1em}
\end{tabular}
}
\end{table*}

\begin{table*}[t]
\caption{Comprehensive comparison of different methods in terms of both average performance and resources utilization. $\uparrow$ means the higher the better, while $\downarrow$ means the lower the better.}
\renewcommand{\figurename}{Fig.}
\centering
\label{tab:performance_resource}
\resizebox{0.9\textwidth}{!}{
\begin{tabular}{c|ccccc|ccccc}
\specialrule{0.2em}{0.1em}{0.1em}
                         & \multicolumn{5}{c|}{Average Performance $\uparrow$}                                                                                                                                                                                                                                   & \multicolumn{5}{c}{Average Resource $\downarrow$}                                                                                                                                                                                                                                    \\ \cline{2-11} 
\multirow{-2}{*}{Methods} & \begin{tabular}[c]{@{}c@{}}Acc.\\ (\%)\end{tabular} & \begin{tabular}[c]{@{}c@{}}Pre.\\ (\%)\end{tabular} & \begin{tabular}[c]{@{}c@{}}Spe.\\ (\%)\end{tabular} & \begin{tabular}[c]{@{}c@{}}Rec.\\ (\%)\end{tabular} & \begin{tabular}[c]{@{}c@{}}F1.\\ (\%)\end{tabular} & \begin{tabular}[c]{@{}c@{}}Footprint\\ (kb)\end{tabular} & MFLOPs                                & \begin{tabular}[c]{@{}c@{}}Latency\\ (ms)\end{tabular} & \begin{tabular}[c]{@{}c@{}}Power\\ (W)\end{tabular} & \begin{tabular}[c]{@{}c@{}}Energy\\ (mJ)\end{tabular} \\ \hline
EEG-ARNN \hspace{0.3ex} \cite{eeg-arnn}                & 61.18                                               & 60.26                                               & 59.10                                               & 62.02                                               & 61.08                                              & {\color[HTML]{9A0000} \textbf{1.30}}                     & 95.46                                 & 85.30                                                  & 6.21                                                & 529.71                                                \\
FBMSNet \hspace{0.3ex} \cite{FBMSNet}                 & 67.40                                               & 72.22                                               & 76.04                                               & 60.46                                               & 65.40                                              & {\color[HTML]{3531FF} \textbf{4.85}}                     & 69.00                                 & 72.62                                                  & 6.02                                                & 437.17                                                \\
LGGNet \hspace{0.3ex} \cite{Lggnet}                  & 70.35                                               & 76.48                                               & 73.80                                               & 66.90                                               & 67.46                                              & 89.49                                                    & 55.11                                 & 72.25                                                  & {\color[HTML]{3531FF} \textbf{5.04}}                & 364.14                                                \\
SSGCNet  \hspace{0.3ex} \cite{ssgcnet}                 & {\color[HTML]{3531FF} \textbf{71.53}}               & {\color[HTML]{3531FF} \textbf{76.60}}               & {\color[HTML]{3531FF} \textbf{76.12}}               & 63.66                                               & 65.11                                              & 62.47                                                    & 33.92                                 & 63.05                                                  & 5.74                                                & 361.90                                                \\
CAGNN \hspace{0.3ex} \cite{cagnn}                    & 66.79                                               & 73.19                                               & 62.28                                               & 66.79                                               & 63.93                                              & 196.82                                                   & 29.91                                 & 40.08                                                  & 5.60                                                & 224.44                                                \\
TSception \hspace{0.3ex} \cite{tsception}                & 69.06                                               & 73.60                                               & 65.74                                               & {\color[HTML]{3531FF} \textbf{72.37}}               & {\color[HTML]{3531FF} \textbf{68.12}}              & 137.98                                                   & {\color[HTML]{3531FF} \textbf{27.71}} & {\color[HTML]{3531FF} \textbf{35.79}}                  & 5.85                                                & {\color[HTML]{3531FF} \textbf{209.37}}                \\ \hline
Ours                     & {\color[HTML]{9A0000} \textbf{77.76}}               & {\color[HTML]{9A0000} \textbf{79.97}}               & {\color[HTML]{9A0000} \textbf{79.26}}               & {\color[HTML]{9A0000} \textbf{76.26}}               & {\color[HTML]{9A0000} \textbf{77.79}}              & 81.24                                                    & {\color[HTML]{9A0000} \textbf{26.26}} & {\color[HTML]{9A0000} \textbf{31.11}}                  & {\color[HTML]{9A0000} \textbf{4.84}}                & {\color[HTML]{9A0000} \textbf{150.57}}                \\ \specialrule{0.2em}{0.1em}{0.1em}
\end{tabular}
}
\end{table*}

\subsection{Comparison with State-of-the-Art Methods}
To evaluate the driver drowsiness monitoring performance of our LDGCN, we compared it with six state-of-the-art methods, including EEG-ARNN \cite{eeg-arnn}, FBMSNet \cite{FBMSNet}, LGGNet \cite{Lggnet}, SSGCNet \cite{ssgcnet}, CAGNN \cite{cagnn}, and TSception \cite{tsception}. Table \ref{tab:perfomance_11} presents the accuracy, precision, and specificity for monitoring drowsiness across 11 different drivers. It is evident that our LDGCN demonstrates robustness across individuals and shows minimal fluctuations due to individual differences.

Furthermore, we comprehensive compare the average recognition performance and resource utilization of different methods, as shown in Table \ref{tab:performance_resource}. For the average performance comparison, our LDGCN achieves the best results in all the five metrics, which are with a 5.79\% accuracy improvement, a 3.37\% precision improvement, and a 3.14\% specificity improvement compared to the second-ranked SSGCNet. Moreover, it also achieves a 3.87\% recall improvement and a 9.69\% F1 Score remarkable improvement compared to the third-ranked TSception. For the resource utilization comparison, our LDGCN outperforms the second-ranked model, reducing Mflops by 1.45, latency by 4.68ms, power by 0.20W, and energy by 58.80mJ. This obviously exhibits the effectiveness of our model in lightweight. Due to the increase in the number of parameters in our graph features comparing to the one-dimensional temporal EEG signal data, our model's footprint is slightly higher than the best-ranked EEG-ARNN. However, with continuous advancements in storage capacity technology, 81.24KB is significantly less than 1MB, so it can render the effect negligible.

\subsection{Ablation Study}
\subsubsection{Impact of the BDSAG}
\begin{figure*}
  \captionsetup{justification=raggedright,singlelinecheck=false}
  \centering
  \renewcommand{\figurename}{Fig.}
  \includegraphics[width=0.99\textwidth]{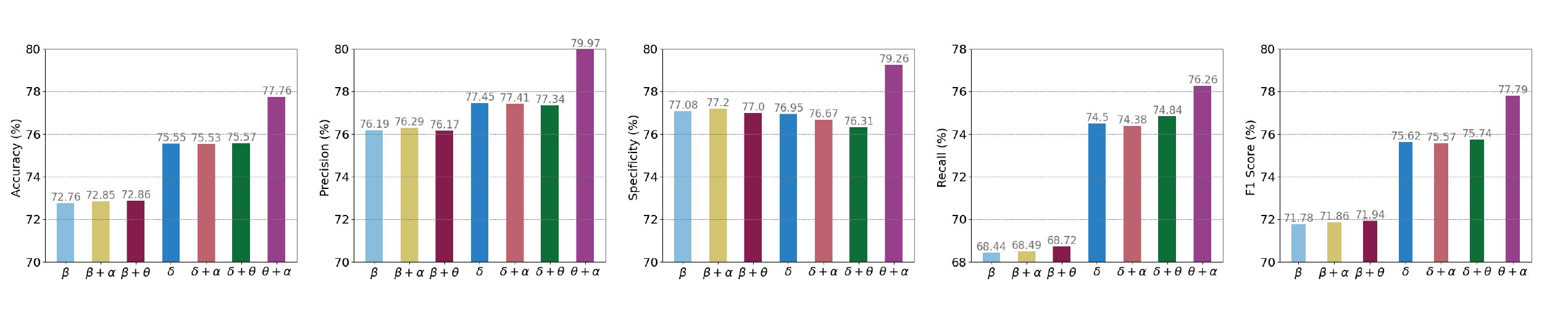}
  \caption{The visual results of five performance metrics of adjacency graphs constructed in different frequency bands.}
  \label{fig:BDSAG_compare}
\end{figure*}

To demonstrate the significant impact of our proposed Baseline Drowsiness Status Adjacency Graph (BDSAG) on drowsiness monitoring performance, we present the results of an ablation study in Figure \ref{fig:BDSAG_compare}. Specifically, instead of using BDSAG, we extracted EEG signals from different frequency bands and constructed new adjacency graphs. Observations reveal that the adjacency graphs created from drowsiness-independent frequency bands, such as $\delta$ and $\beta$, show significant degradation across all five performance metrics. This decline in performance is attributed to the inadequacy of the $\delta$ and $\beta$ frequency bands in effectively capturing features related to the driver's drowsiness status. As a result, the expressive ability of these constructed adjacency graphs is limited, negatively impacting the recognition performance of the LDGCN model. This outcome aligns precisely with the findings of Lal \textit{et al.} \cite{i12}, confirming the close relationship between the $\theta$ and $\alpha$ frequency bands in relation to drowsiness status.

\subsubsection{Impact of the AGM}
To validate the effectiveness of our proposed Augmented Graph-level Module (AGM), we assessed the performance of various data augmentation modules in drowsiness monitoring. As shown in Table \ref{tab:Augmentations_data}, when employing AGM in the LDGCN, all five performance evaluation metrics show notable improvements. Conversely, the use of other popular data augmentation techniques results in performance decreases to varying degrees. The primary reason for this discrepancy lies in the inapplicability of these methods to our constructed Baseline Drowsiness Status Adjacency Graph (BDSAG). In other words, these methods compromise the integrity of the features, leading to significant errors in feature representation, which ultimately impacts the overall performance metrics negatively.

\begin{table}[h]
\caption{Performance comparison of different data augmentation modules. The best data are marked in red.}
\renewcommand{\figurename}{Fig.}
\centering
\label{tab:Augmentations_data}
\resizebox{0.49\textwidth}{!}{
\begin{tabular}{c|ccccc}
\hline
Augmentation modules    & \begin{tabular}[c]{@{}c@{}}Acc.\\ (\%)\end{tabular} & \begin{tabular}[c]{@{}c@{}}Pre.\\ (\%)\end{tabular} & \begin{tabular}[c]{@{}c@{}}Spe.\\ (\%)\end{tabular} & \begin{tabular}[c]{@{}c@{}}Rec.\\ (\%)\end{tabular} & \begin{tabular}[c]{@{}c@{}}F1.\\ (\%)\end{tabular} \\ \hline
w/o AGM   & 76.79                                               & 78.86                                               & 78.36                                               & 74.92                                               & 76.66                                              \\
GN \hspace{0.3ex} \cite{Gaussiannoise}         & 71.63                                               & 74.16                                               & 73.94                                               & 69.32                                               & 71.36                                              \\
GraphSMOTE \hspace{0.3ex} \cite{Graphsmote}  & 72.26                                               & 75.02                                               & 74.46                                               & 70.05                                               & 72.06                                              \\
SimPSI \hspace{0.3ex} \cite{SimPSI}     & 71.32                                               & 72.26                                               & 72.68                                               & 69.97                                               & 70.87                                              \\ \hline
AGM       & {\color[HTML]{9A0000} \textbf{77.76}}               & {\color[HTML]{9A0000} \textbf{79.97}}               & {\color[HTML]{9A0000} \textbf{79.26}}               & {\color[HTML]{9A0000} \textbf{76.26}}               & {\color[HTML]{9A0000} \textbf{77.79}}              \\ \hline
\end{tabular}
}
\end{table}

\subsubsection{Impact of the APO}
To validate the effectiveness of our Adaptive Pruning Optimization (APO) strategy, we compared it with other popular pruning methods. As shown in Table \ref{tab:pruning_compare}, employing other pruning methods generally results in a decrease in performance metrics to varying degrees. In contrast, utilizing our APO, the LDGCN demonstrates superior performance in accuracy, precision, and specificity. Additionally, it achieves a significant reduction in model testing time, halving it to approximately 44.44 ms. Notably, compared to ADMM, our APO strategy enhances accuracy by 0.9\%, precision by 2.49\%, and specificity by 3.45\%. All data indicate that our APO not only prunes more parameters effectively but also reduces latency by 5.31 ms. Overall, APO excels in optimizing both channels and neurons. Its filtering process can operate across a significant number of channels and neurons, accomplishing lossless pruning and thus substantially reducing latency during testing.

\begin{table}[h]
\caption{Performance comparison of different pruning methods. The best data are marked in red.}
\renewcommand{\figurename}{Fig.}
\centering
\label{tab:pruning_compare}
\resizebox{0.49\textwidth}{!}{
\begin{tabular}{c|ccccc}
\specialrule{0.2em}{0.1em}{0.1em}
Pruning strategies  & \begin{tabular}[c]{@{}c@{}}Acc.\\ (\%)\end{tabular} & \begin{tabular}[c]{@{}c@{}}Pre.\\ (\%)\end{tabular} & \begin{tabular}[c]{@{}c@{}}Spe.\\ (\%)\end{tabular} & \begin{tabular}[c]{@{}c@{}}Pruning\\ Params\end{tabular} & \begin{tabular}[c]{@{}c@{}}Latency\\ (ms)\end{tabular} \\ \hline
w/o APO & 76.96                                               & 79.38                                               & 79.21                                               & 0                                                        & 75.55                                                  \\
L1 \hspace{0.3ex} \cite{r4}      & 70.24                                               & 67.55                                               & 61.85                                               & 8346                                                     & 64.30                                                  \\
SlimNet \hspace{0.3ex} \cite{slimnet} & 72.49                                               & 72.89                                               & 78.73                                               & 9685                                                     & 47.46                                                  \\
WMB \hspace{0.3ex} \cite{wmb}     & 73.68                                               & 76.58                                               & 77.11                                               & 792                                                      & 66.06                                                  \\
ADMM  \hspace{0.3ex} \cite{ssgcnet}   & 76.86                                               & 77.48                                               & 75.81                                               & 13365                                                    & 36.42                                                  \\ \hline
APO     & {\color[HTML]{9A0000} \textbf{77.76}}               & {\color[HTML]{9A0000} \textbf{79.97}}               & {\color[HTML]{9A0000} \textbf{79.26}}               & {\color[HTML]{9A0000} \textbf{17502}}                    & {\color[HTML]{9A0000} \textbf{31.11}}                  \\ \specialrule{0.2em}{0.1em}{0.1em}
\end{tabular}
}
\end{table}

\begin{figure}
  \captionsetup{justification=raggedright,singlelinecheck=false}
  \centering
  \renewcommand{\figurename}{Fig.}
  \includegraphics[width=0.35\textwidth]{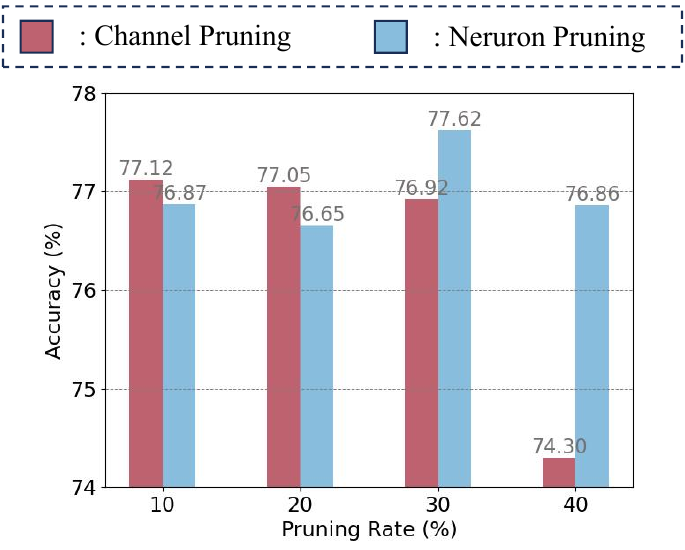}
  \caption{The accuracy comparison of using channel pruning and neuron pruning optimization strategies, separately.}
  \label{fig:pruning_1}
\end{figure}

\begin{figure}
  \captionsetup{justification=raggedright,singlelinecheck=false}
  \centering
  \renewcommand{\figurename}{Fig.}
  \includegraphics[width=0.49\textwidth]{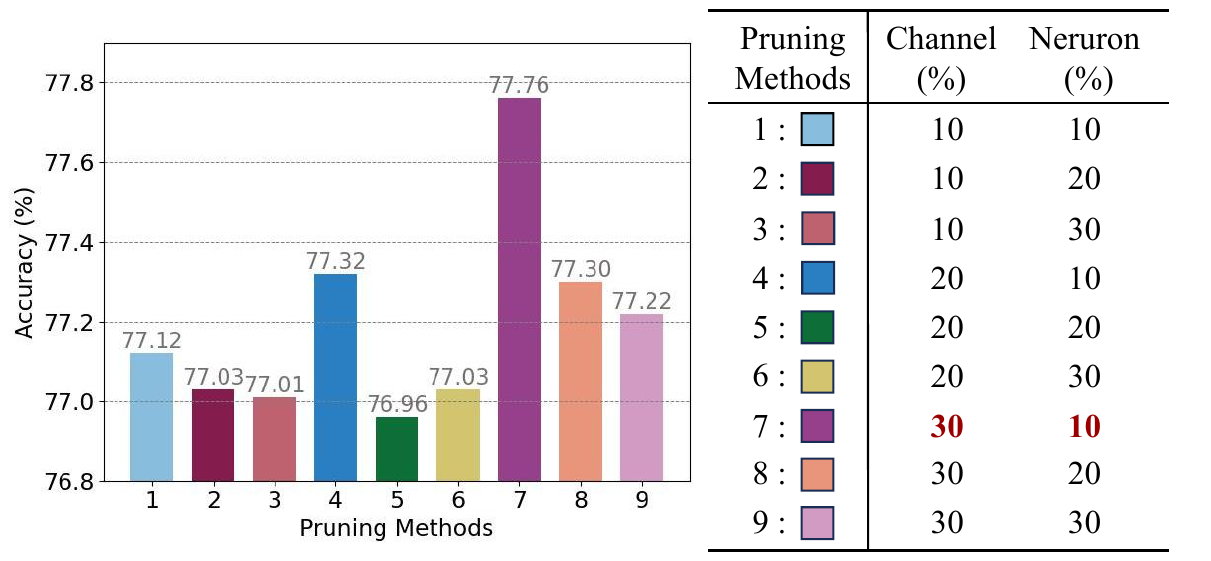}
  \caption{The accuracy comparison of using combined channel pruning and neuron pruning optimization.}
  \label{fig:pruning_2}
\end{figure}

Additionally, we optimized channel and neuron pruning separately with varying ratios. As illustrated in Figure \ref{fig:pruning_1}, LDGCN achieves the highest accuracy when channels are pruned by only 10\% and neurons by 30\%. Based on these findings, it can be hypothesized that further accuracy improvements could be attained by simultaneously pruning 10\% of channels and 30\% of neurons. Consequently, we conducted combined pruning experiments, as depicted in Figure \ref{fig:pruning_2}. When applying both 10\% channel pruning and 30\% neuron pruning simultaneously, the model not only achieves the best accuracy but also outperforms other combined pruning strategies. This result validates the effectiveness of our proposed channel and neuron pruning optimization, demonstrating its ability to achieve a trade-off between performance and resource utilization.

\subsubsection{Impact of the LDGCN Structure}
We present the ablation results to assess the effectiveness of each layer in the LDGCN model, as shown in Table \ref{tab:netmodule}. The specific experimental groups in Table \ref{tab:netmodule} are as following.

(1) Conv. layer designs ‘1-2’: We evaluate the impact of using a single layer of graph convolution. This includes pointwise convolution and depthwise convolution with a single layer ($\mathrm{PDwise} \times 1$) , as well as a fully connected layer with a single layer ($\mathrm{FC} \times 1$). Compared to the full model, it exhibits a 2.17\% and 3.64\% decrease in accuracy, respectively. This validates the effectiveness of our lightweight dual graph convolution structure for feature learning. 

(2) Conv. layer designs ‘3-4’: We separately evaluate the impact of pointwise convolution and depthwise convolution on drowsiness monitoring. The results indicate that both pointwise convolution across channels and depthwise convolution within channels are highly effective. This validates that even after the convolution operation, our graph features maintain strong integrity, which is contributing to their expressive power. 

(3) Conv. layer design ‘5’: We evaluate the effectiveness of partial convolution in the fusion layer. After removing partial convolution, the accuracy drops by 2.1\%. This validates that partial convolution not only selectively controls information flow, making the network more compact, but also exhibit the impact of feature transfer and interaction.

\begin{table}[h]
\caption{Ablation studies of each important designed layer in LDGCN}
\renewcommand{\figurename}{Fig.}
\centering
\label{tab:netmodule}
\resizebox{0.49\textwidth}{!}{
\begin{tabular}{cc|ccccc}
\specialrule{0.2em}{0.1em}{0.1em}
\multicolumn{2}{c|}{Different Conv. Layer Designs}               & \begin{tabular}[c]{@{}c@{}}Acc.\\ (\%)\end{tabular} & \begin{tabular}[c]{@{}c@{}}Pre.\\ (\%)\end{tabular} & \begin{tabular}[c]{@{}c@{}}Spe.\\ (\%)\end{tabular} & \begin{tabular}[c]{@{}c@{}}Rec.\\ (\%)\end{tabular} & \begin{tabular}[c]{@{}c@{}}F1.\\ (\%)\end{tabular} \\ \hline
\multicolumn{1}{c|}{1} & PDwise $\times$ 1          & 75.59                                 & 76.42                                 & 74.94                                 & 76.17                                 & 76.47                                 \\
\multicolumn{1}{c|}{2} & FC $\times$ 1           & 74.12                                 & 74.39                                 & 73.30                                 & 74.94                                 & 74.41                                 \\
\multicolumn{1}{c|}{3} & w/o Pointwise    & 73.43                                 & 74.91                                 & 71.74                                 & 75.11                                 & 74.11                                 \\
\multicolumn{1}{c|}{4} & w/o Depthwise    & 73.92                                 & 77.28                                 & 75.63                                 & 72.20                                 & 73.83                                 \\
\multicolumn{1}{c|}{5} & w/o PartialConv & 75.66                                 & 77.65                                 & 76.57                                 & 74.74                                 & 75.79                                 \\ \hline
\multicolumn{2}{c|}{Ours (full model)}               & {\color[HTML]{9A0000} \textbf{77.76}} & {\color[HTML]{9A0000} \textbf{79.97}} & {\color[HTML]{9A0000} \textbf{79.26}} & {\color[HTML]{9A0000} \textbf{76.26}} & {\color[HTML]{9A0000} \textbf{77.79}} \\ \specialrule{0.2em}{0.1em}{0.1em}
\end{tabular}
}
\end{table}

\section{CONCLUSIONS} \label{sec5:conclusion}
In order to achieve monitor driver drowsiness status in real-time on resource-limited edge-end devices, this work proposes a Lightweight Dual GCN (LDGCN) method based on single-channel driver drowsiness EEG signal. The LDGCN is guaranteed to be lightweight while applying a dual graph convolution structure to enhance the robustness of the model. To our best knowledge, we are the first to consider neurophysiological knowledge to construct a Baseline Drowsiness Status Adjacency Graph (BDASG) with interpretable features, which can effectively represent the EEG signals associated with drowsiness. To improve the feature representation, we introduce an Augmented Graph-level Module (AGM) that guarantees the integrity of BDSAG features and extracts global and local view features. Moreover, to reduce the computational complexity of the model and deploy on the fourth generation Raspberry Pi, we optimize the channels and neurons with Adaptive Pruning Optimization (APO), which reduces the inference time by half. Compared to the six state-of-the-art models, our LDGCN exhibits higher recognition performance and lower resource consumption on a benchmark dataset. Furthermore, comprehensive ablation studies on the proposed BDASG, AGM, APO, and the model structure have validated that our LDGCN strikes a balance between performance and resources, making it a valuable approach for application in edge-end AI devices for assisting driver safety.

\section*{ACKNOWLEDGMENT}
This work is supported by the First Batch of Industry-University Cooperation Collaborative Education Project Funded by the Ministry of Education of the People’s Republic of China, 2021, Project No.202101071001, Minjiang University School-Level Scientific Research Project Funding, Projects No. MYK17021, MYK18033, MYK21011, Minjiang University Introduced Talents Scientific Research Start-up Fund, Projects No. MJY21030, Digital Media Art, Key Laboratory of Sichuan Province, Sichuan Conservatory of Music, Project No. 21DMAKL01, Digital Media Art, Key Laboratory of Sichuan Province, Sichuan Conservatory of Music, Project No. 23DMAKL02, The Ministry of Education’s first batch of industry-university cooperation collaborative education projects in 2024 (project number: 231104497282106), Fujian Province Software Industry Technology Innovation Key Research and Industrialization Project (Second Batch), "AI Intelligent Modeling Platform Based on Deep Learning Technology" Project, and Young and Middle-aged Teacher Education Research Project of Fujian Province (Science and Technology, No. JAT220829.

\normalem
\bibliographystyle{ieeetr}
\bibliography{ref}
\end{document}